\documentclass[10pt,a4paper]{article}
\usepackage{amsmath,amssymb}
\usepackage{graphicx}
\graphicspath{{figures/}}
\usepackage{hyperref}
\usepackage[margin=1in]{geometry}
\usepackage{float}
\usepackage[utf8]{inputenc}
\usepackage{authblk}

\title{Spall Failure of Coral Skeleton beneath Gas-Laden Canopies:\\
An Idealized Blast-Fishing Model}

\author[1,2]{Sandy H. S. Herho}
\author[3]{Agus W. Jatmiko}
\author[4,5]{Rizki D. Permana}
\author[5]{Iwan P. Anwar}
\author[2,6]{Alfita P. Handayani}
\author[5]{Faruq Khadami}
\author[5]{Karina A. Sujatmiko}
\author[7]{Rusmawan Suwarman}
\author[1]{Deny J. Puradimaja}
\author[1,*]{Dasapta E. Irawan}

\affil[1]{Applied Geology Research Group, Bandung Institute of Technology, Bandung, West Java, Indonesia}
\affil[2]{Center for Agrarian Studies, Bandung Institute of Technology, Bandung, West Java 40132, Indonesia}
\affil[3]{Headquarters of the Indonesian Armed Forces (Mabes TNI),
Cilangkap, East Jakarta 13870, Indonesia}
\affil[4]{Marine Sciences and Technology Research Group, Sumatera Institute of Technology, Southern Lampung, Lampung 35365, Indonesia}
\affil[5]{Applied and Environmental Oceanography Research Group, Bandung Institute of Technology, Bandung, West Java, Indonesia}
\affil[6]{Spatial System and Cadaster Research Group, Bandung Institute of Technology, Bandung, West Java 40132, Indonesia}
\affil[7]{Atmospheric Science Research Group, Bandung Institute of Technology, Bandung, West Java 40132, Indonesia}
\affil[*]{Corresponding author. E-mail: dasaptaerwin@itb.ac.id}

\date{}

\begin{document}
\maketitle

\begin{abstract}
\noindent
Improvised explosives used for fishing on shallow Indonesian reefs shatter coral skeleton, yet the damage they cause has been described largely through empirical radii and ecological surveys. We formulate an idealized model of how free gas held within a coral canopy modifies the shock loading that such a charge delivers to skeletal plates. The canopy is treated as a relaxed bubbly mixture whose shock impedance follows from the conservation of mass and momentum, and the reef is represented as a layered column of water, canopy, skeletal plate, and canopy struck at normal incidence. Three closed-form results emerge. Above a crossover pressure set by the void fraction and the stiffness of seawater, the canopy becomes nearly transparent to the shock. The impulse transmitted through any lossless layered stack is independent of the canopy, so gas redistributes the pulse in time without changing its total push. A plate carries tension after reflection from its lower face only when the canopy impedance falls below a threshold fixed by the plate thickness and the pulse duration, which defines a critical thickness. For a one-kilogram charge directly overhead, a gas-rich canopy more than triples the standoff at which a twelve-centimeter plate spalls while shortening the standoff at which it is crushed. A prescribed daily cycle of photosynthetic gas makes the same charge markedly more damaging at noon than at night. Bubble dynamics show that the canopy does not reach equilibrium within the pulse, so the results are best read as upper bounds.
\end{abstract}


\section{Introduction}

Fishing with improvised explosives has degraded coral reefs across Southeast Asia for several decades, and Indonesia has been among the most affected regions \cite{PetSoede1999,Edinger1998,HamptonSmith2021}. The same coastal waters are under pressure from industrial and land-based sources that degrade water clarity and promote algal blooms \cite{Herho2026morowali,Anwar2026balikpapan}. A typical charge consists of fertilizer and fuel packed into a glass bottle or a drum, thrown from a small boat, and detonated above or within the reef framework. The blast kills or stuns fish through barotrauma and simultaneously shatters the carbonate skeleton of nearby colonies. Field studies in Komodo and Bunaken National Parks found no significant natural recovery in blast-created rubble fields monitored over several years, largely because mobile rubble abrades and buries new recruits \cite{Fox2003}. Craters produced by isolated blasts recovered over roughly five years, whereas extensively bombed areas showed no recovery over six years despite adequate larval supply \cite{FoxCaldwell2006}. Earlier Philippine and Indonesian surveys reached similar conclusions about recovery rates after destructive fishing \cite{McManus1997,Edinger1998}, and population-level models have treated repeated blasting as a disturbance term acting on coral cover \cite{Saila1993}.

The physical side of the problem has received far less attention than its ecological and economic consequences. Economic analyses have quantified the private gains and social losses of blast fishing on Indonesian reefs \cite{PetSoede1999}, and a recent global review compiled its causes, extent, and management responses \cite{HamptonSmith2021}. Acoustic work has focused on detection: blast signatures recorded in the field can be separated from ambient reef noise and located by triangulation, which supports enforcement \cite{Woodman2003,Showen2018}. The mechanics of skeletal failure under blast loading, and the role of the reef environment in shaping that loading, have not been formulated from first principles to our knowledge. Damage is usually summarized as an empirical destructive radius that depends only on charge size.

The incident loading itself is well characterized. A detonation in open water produces a shock whose peak pressure and exponential decay constant follow similitude laws in the scaled range $W^{1/3}/R$, where $W$ is the charge mass and $R$ the range \cite{Cole1948}. Measurements in shallow water for charges between 0.1 and 6~kg TNT equivalent agree well with the similitude law for peak pressure \cite{SolowayDahl2014}, which covers the size range of improvised fishing charges. The later shape of the pulse and the pressure field of the explosion bubble depend more strongly on charge size and depth \cite{GeersHunter2002,HunterGeers2004}. Near the free surface, the surface-reflected rarefaction produces bulk cavitation of the upper water column, which is also the zone of greatest mortality for fish with swim bladders \cite{Cole1948,Saila1993}.

The medium that this shock enters above a reef is not ordinary seawater. Coral canopies modify oscillatory flow, mass transfer, and turbulence in ways that have been studied extensively \cite{Monismith2007,Lowe2005,Nepf2012}, and idealized models of wave attenuation through coastal vegetation show how canopy drag and geometry control the energy transmitted across such layers \cite{Herho2026wave}. Photosynthetic canopies can also hold free gas. In a seagrass meadow, the diel cycle of acoustic transmission tracks oxygen production closely enough that bubble-mediated attenuation has been used as a proxy for primary productivity \cite{Felisberto2015}. Even small volume fractions of gas change the acoustics of water drastically. The low-frequency sound speed of the mixture falls to a small fraction of that of either phase \cite{Wood1930}, and the theory of linear waves in bubbly liquids, including dispersion and attenuation near bubble resonance, is well established \cite{Carstensen1947,CommanderProsperetti1989,vanWijngaarden1972}. Shock waves in bubbly liquids behave differently again: their speed depends on both void fraction and shock strength, and their structure is controlled by bubble oscillation and relative motion \cite{CampbellPitcher1958,Noordzij1974,Brennen2013}.

This combination suggests a mechanism that has not been examined. A skeletal plate immersed in a gas-laden canopy is bounded on its lower face by a medium of low impedance. When a compressive pulse transmitted into the plate reaches that face, it reflects with the sign of a free surface and places the plate in tension, the classical setting for spallation \cite{Grady1988,Antoun2003}. Coral skeleton fails in compression at stresses between about 12 and 81~MPa depending on species and porosity \cite{Chamberlain1978}, brittle porous solids are generally several times weaker in tension than in compression \cite{Meyers1994}, and colony dislodgement and breakage under hydrodynamic loading already shape reef assemblages \cite{MadinConnolly2006}. Whether canopy gas can convert compressive loading into tensile failure, and over what ranges of charge, standoff, void fraction, and plate thickness, is a question of wave mechanics that can be posed exactly in an idealized setting.

We formulate that setting here. The canopy is represented as a relaxed bubbly mixture, the reef as a one-dimensional column at normal incidence, and the charge by the similitude pulse. The analysis yields closed forms for the canopy shock impedance, the crossover overpressure, the transmitted impulse, the spall onset criterion, and the thickness of the spall scab. These are evaluated with three independent layered solvers, tested against Keller-Miksis bubble dynamics, and illustrated with two-dimensional linear acoustics over a branching thicket and a tabular plate. The approach follows a series of idealized open-source solvers for geophysical and nonlinear wave problems, in which transparent numerics and exact test cases are used to isolate mechanisms, including stratified shear instability \cite{Herho2025kh2d}, dam-break bores \cite{Irawan2026amerta}, and nonlinear dispersive waves \cite{Irawan2026kdv,Herho2026nlse}. The model is deliberately idealized. It contains no field calibration and is intended to identify the mechanism, its governing dimensionless groups, and the conditions under which it can matter.

\section{Methods}

\subsection{Model Description}

\begin{figure}[H]
\centering
\includegraphics[width=0.9\textwidth]{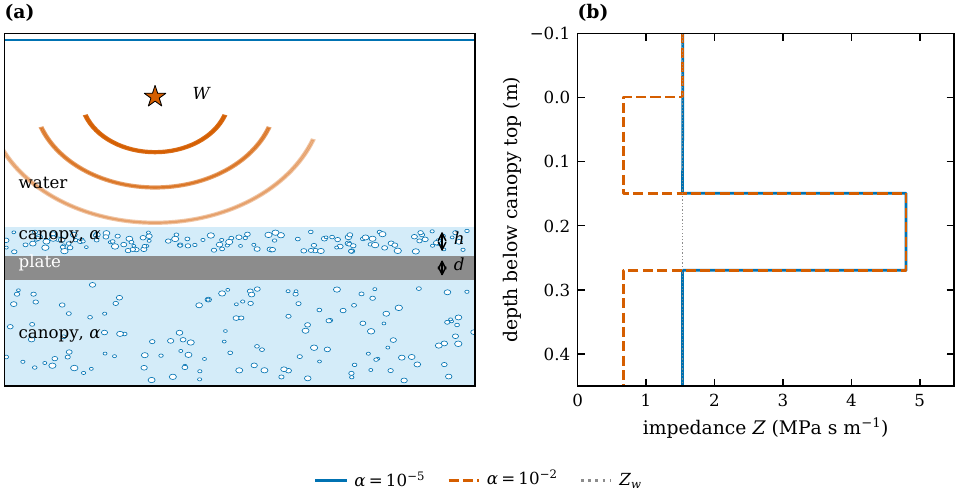}
\caption{Model configuration. (a) A charge in the water column, drawn as an orange star with vermilion arcs for the outgoing front, above a gas-laden canopy of thickness $h$ over a skeletal plate of thickness $d$, with canopy below the plate. Water and canopy are shaded pale blue, free gas is drawn as open circles, and the plate is grey. (b) Impedance against depth for a gas-free canopy, $\alpha=10^{-5}$, as a solid blue staircase, and for $\alpha=10^{-2}$ as a dashed vermilion staircase, both evaluated with the secant impedance at 5~MPa. The vertical grey dotted line marks the impedance of seawater. The plate is the only layer stiffer than seawater, so at sufficiently low canopy impedance its lower face reflects compressive waves with the sign appropriate to a free surface.}
\label{fig:schematic}
\end{figure}

The configuration is shown in Fig.~\ref{fig:schematic}. A charge of TNT-equivalent mass $W$ detonates at vertical standoff $R$ above a tabular skeletal plate of thickness $d$. The plate is covered by a canopy layer of thickness $h$ and underlain by canopy that extends to depth. Both canopy layers consist of seawater carrying a volume fraction $\alpha$ of free gas. The model is built in four steps: a continuum description of the gas-laden canopy and its response to a shock, the incident loading, the propagation of that loading through the layered column into the plate, and two auxiliary descriptions that test the relaxed closure and illustrate the pattern of loading over an irregular reef.

We begin with the canopy. Let bubbles of radius $a$ be separated by a mean distance $\ell$, and let $\lambda$ be the shortest wavelength of interest. When $a\ll\ell\ll\lambda$, averaging over volumes that contain many bubbles but are small compared with $\lambda$ defines a mixture continuum with density $\rho$, velocity $u$, and pressure $p$ \cite{vanWijngaarden1972,Brennen2013}. With $\rho_g\ll\rho_l$, the mixture density is
\begin{equation}
\rho=(1-\alpha)\rho_l+\alpha\rho_g\simeq(1-\alpha)\rho_l .
\label{eq:rhomix}
\end{equation}
Because the charge is directly overhead and the radius of curvature of the front, which equals $R$, is several times the column thickness $h+d$ at the standoffs of interest, the motion is treated as planar along the vertical coordinate $z$, positive downward. The neglected spherical spreading changes amplitudes across the column by a fraction of order $(h+d)/R$, which is below fifteen percent for $R>2$~m. Conservation of mass and momentum for a fixed interval $[z_1,z_2]$ read
\begin{equation}
\frac{d}{dt}\int_{z_1}^{z_2}\rho\,dz=-\big[\rho u\big]_{z_1}^{z_2},\qquad \frac{d}{dt}\int_{z_1}^{z_2}\rho u\,dz=-\big[\rho u^2+p\big]_{z_1}^{z_2},
\label{eq:integral}
\end{equation}
and, for smooth fields and an arbitrary interval, reduce to
\begin{equation}
\partial_t\rho+\partial_z(\rho u)=0,\qquad \partial_t(\rho u)+\partial_z(\rho u^2+p)=0 .
\label{eq:euler}
\end{equation}
Viscous stresses and gravity are omitted because the shock rise time and pulse duration, of order $10^{-4}$~s, are short compared with viscous diffusion times over the column and with the gravitational time scale $\sqrt{h/g}$ \cite{Meyers1994}. The system is closed by a relation between specific volume $v=1/\rho$ and pressure. Two assumptions define the relaxed mixture. The phases move with one velocity, which neglects bubble slip, and the gas pressure equals the liquid pressure, which neglects bubble inertia and surface tension on the scale of the wave \cite{vanWijngaarden1972}. Their validity is tested below with the Keller-Miksis equation.

Neglecting gas mass, the specific volume per unit mass of mixture is the sum of a liquid and a gas contribution. At the ambient absolute pressure $p_0$ these are
\begin{equation}
v_{l0}\equiv\frac{1}{\rho_l},\qquad v_{g0}\equiv\frac{\alpha}{(1-\alpha)\rho_l},\qquad v_0\equiv v_{l0}+v_{g0}=\frac{1}{(1-\alpha)\rho_l}.
\label{eq:v0}
\end{equation}
The liquid obeys the definition of its isentropic bulk modulus, $dv_l/v_l=-dp/K_l$ with $K_l=\rho_lc_l^2$. Integrating to first order in $\Delta p/K_l$ gives $v_l=v_{l0}(1-\Delta p/K_l)$, which is adequate because $\Delta p/K_l$ remains near two percent even at 50~MPa. The gas is compressed along the polytrope $p\,v_g^{\kappa}=p_0v_{g0}^{\kappa}$. The exponent is set by the thermal P\'{e}clet number $\mathrm{Pe}\equiv\omega R_0^2/D_{\mathrm{th}}$, where $\omega$ is the angular frequency of bubble motion, $R_0$ the bubble radius, and $D_{\mathrm{th}}$ the thermal diffusivity of the gas. For a 0.5~mm bubble oscillating near its natural frequency, $\mathrm{Pe}$ is of order $6\times10^2$, so heat does not diffuse out of the bubble within a cycle and the compression is close to adiabatic, $\kappa\simeq\gamma=1.4$ \cite{PlessetProsperetti1977}. The change of specific volume produced by an overpressure $\Delta p$ is therefore
\begin{equation}
v_0-v_1=\frac{v_{l0}\,\Delta p}{K_l}+v_{g0}\left[1-\left(\frac{p_0}{p_0+\Delta p}\right)^{1/\kappa}\right].
\label{eq:dv}
\end{equation}

A shock is a moving discontinuity across which Eqs.~\eqref{eq:integral} still hold. Let the front move with speed $U$ into mixture at rest in state $(v_0,p_0,u_0=0)$ and leave behind state $(v_1,p_0+\Delta p,u_1)$. Applying Eqs.~\eqref{eq:integral} to an interval that contains the front and shrinks with it gives the Rankine-Hugoniot conditions, the same construction that governs hydraulic bores in shallow water \cite{Meyers1994,Irawan2026amerta}. In the frame of the front, the mass flux $m$ and the momentum balance are
\begin{equation}
m=\frac{U}{v_0}=\frac{U-u_1}{v_1},\qquad \Delta p=m\,u_1 .
\label{eq:rh}
\end{equation}
The first relation gives $u_1=U(v_0-v_1)/v_0$. Substituting into the second yields $\Delta p=U^2(v_0-v_1)/v_0^2$, and therefore
\begin{equation}
U^2=\frac{v_0^2\,\Delta p}{v_0-v_1},\qquad Z_c\equiv\frac{\Delta p}{u_1}=\frac{U}{v_0}=\rho_0U .
\label{eq:rayleigh}
\end{equation}
The first expression is the Rayleigh line of the mixture, and $Z_c$ is its secant, or shock, impedance, the ratio of pressure jump to particle-velocity jump. The energy balance is not needed to close Eq.~\eqref{eq:rayleigh} because the constitutive relation \eqref{eq:dv} is barotropic. Physically, the energy that a barotropic shock does not account for is carried into bubble oscillation and eventually into heat behind the front \cite{CampbellPitcher1958,Noordzij1974}.

Two limits organize the behavior of Eq.~\eqref{eq:rayleigh}. For small overpressure the bracket in Eq.~\eqref{eq:dv} expands as $\Delta p/(\kappa p_0)-(1+\kappa)\Delta p^2/(2\kappa^2p_0^2)+O(\Delta p^3)$. The leading term gives
\begin{equation}
\lim_{\Delta p\to0}U^2=c_W^2,\qquad \frac{1}{\rho_mc_W^2}=\frac{1-\alpha}{K_l}+\frac{\alpha}{\kappa p_0},\qquad \rho_m=(1-\alpha)\rho_l ,
\label{eq:wood}
\end{equation}
which is the classical low-frequency mixture relation, and $c_W$ is referred to below as the Wood speed \cite{Wood1930,Carstensen1947}. The quadratic term shows that $U$ departs from $c_W$ linearly in $\Delta p$. The physical content of Eq.~\eqref{eq:wood} is that the mixture takes its inertia almost entirely from the liquid but its compressibility almost entirely from the gas, which is why a void fraction of $10^{-3}$ can reduce the sound speed to about one third of $c_l$ \cite{CommanderProsperetti1989}. For large overpressure the gas contribution in Eq.~\eqref{eq:dv} cannot exceed $v_{g0}$, because a bubble cannot be compressed below zero volume, whereas the liquid contribution grows without bound. The two contributions are equal at the crossover overpressure
\begin{equation}
p^*\equiv\frac{v_{g0}}{v_{l0}}\,K_l=\frac{\alpha K_l}{1-\alpha}.
\label{eq:pstar}
\end{equation}
For $\Delta p\gg p^*$ the liquid carries most of the compliance and the canopy is nearly transparent to the shock. At $\Delta p=p^*$ the gas term equals $v_{g0}(1-\varepsilon)$ with $\varepsilon\equiv(1+p^*/p_0)^{-1/\kappa}$, so $v_0-v_1=v_{g0}(2-\varepsilon)$. Using $p^*=K_lv_{g0}/v_{l0}$, $v_0=v_{l0}/(1-\alpha)$, and $K_lv_{l0}=c_l^2$ in Eq.~\eqref{eq:rayleigh} gives the exact result
\begin{equation}
U(p^*)=\frac{c_l}{(1-\alpha)\sqrt{2-\varepsilon}},\qquad \varepsilon\equiv\left(1+\frac{p^*}{p_0}\right)^{-1/\kappa},
\label{eq:ustar}
\end{equation}
which tends to $c_l/[\sqrt{2}(1-\alpha)]$ as $p^*/p_0$ grows.

The incident loading is represented through Hopkinson-Cranz scaling. If the energy released is proportional to $W$ and the explosive and water properties are fixed, dimensional analysis requires that pressure depend on range only through the scaled distance $R/W^{1/3}$ and that times scale with $W^{1/3}$ \cite{Cole1948}. Power-law fits within this form give
\begin{equation}
p_i(t)=P\,e^{-t/\theta}H(t),\qquad P=K_P\left(\frac{W^{1/3}}{R}\right)^{A_P},\qquad \theta=K_T\,W^{1/3}\left(\frac{W^{1/3}}{R}\right)^{-A_T},
\label{eq:similitude}
\end{equation}
where $H$ is the Heaviside function and $(K_P,A_P,K_T,A_T)=(52.16~\text{MPa},1.13,92.5~\mu\text{s\,kg}^{-1/3},0.22)$ are the widely tabulated TNT constants \cite{Cole1948,SolowayDahl2014}. An acoustic spherical wave would give $A_P=1$. The larger exponent reflects additional dissipation at the shock front, and the positive $A_T$ reflects the lengthening of the pulse as the front weakens. Improvised ammonium-nitrate charges enter only through an equivalent $W$. The exponential form is accurate for about one decay constant, after which the pressure decays more slowly than Eq.~\eqref{eq:similitude} implies \cite{Cole1948,GeersHunter2002}.

Propagation through the column is treated with linear acoustics in each layer, with the nonlinearity of the canopy retained through its impedance. Linearizing Eqs.~\eqref{eq:euler} about rest in a homogeneous layer $j$ of density $\rho_j$ and sound speed $c_j$ gives
\begin{equation}
\rho_j\,\partial_tu=-\partial_zp,\qquad \partial_tp=-\rho_jc_j^2\,\partial_zu ,
\label{eq:acoustic1d}
\end{equation}
and eliminating $u$ yields the wave equation $\partial_t^2p=c_j^2\partial_z^2p$. Its general solution and the corresponding velocity are
\begin{equation}
p=f\!\left(t-\frac{z}{c_j}\right)+g\!\left(t+\frac{z}{c_j}\right),\qquad u=\frac{1}{Z_j}\left[f\!\left(t-\frac{z}{c_j}\right)-g\!\left(t+\frac{z}{c_j}\right)\right],\qquad Z_j\equiv\rho_jc_j ,
\label{eq:dalembert}
\end{equation}
where $f$ travels downward and $g$ upward \cite{Pierce2019}. At a material interface, Eqs.~\eqref{eq:integral} applied to a vanishing interval that contains the interface require continuity of normal velocity and of pressure. For a wave of amplitude $f$ incident from medium 1 onto medium 2, with reflected amplitude $\mathcal{R}f$ and transmitted amplitude $\mathcal{T}f$, these conditions read $1+\mathcal{R}=\mathcal{T}$ and $(1-\mathcal{R})/Z_1=\mathcal{T}/Z_2$, whose solution is
\begin{equation}
R_{12}=\frac{Z_2-Z_1}{Z_2+Z_1},\qquad T_{12}=\frac{2Z_2}{Z_1+Z_2}.
\label{eq:rt}
\end{equation}
The skeleton is treated as an acoustic medium with longitudinal speed $c_s$ and impedance $Z_s=\rho_sc_s$, so the stress normal to the plate is $\sigma=-p$ and tension corresponds to negative $p$.

The canopy is not linear, and its impedance depends on the strength of the wave it carries. At an interface between water and canopy, the state transmitted into the canopy must lie on the canopy Hugoniot $p=Z_c(p)\,u$, while the state on the water side must lie on the acoustic reflection line $p=2P-Z_wu$. Their intersection in the pressure and particle-velocity plane gives the transmitted pressure $p_t=2Z_c(p_t)P/[Z_w+Z_c(p_t)]$, which is the impedance-matching construction of shock physics \cite{Meyers1994}. We evaluate $Z_c$ at the incident peak $P$ rather than at $p_t$ and hold it fixed during the pulse. This frozen-secant approximation is exact in the acoustic limit, preserves the linear superposition needed for the layered solution, and overstates the canopy impedance below the plate, where the transmitted pulse is weaker than $P$.

Normal incidence is essential to the layered description. For a plane wave $p\propto\exp[\mathrm{i}(\omega t-k_xx-k_zz)]$, continuity of pressure along an interface for all $x$ requires the horizontal wavenumber $k_x\equiv\omega\sin\vartheta/c$ to be the same in every layer. Hence $\sin\vartheta_j/c_j$ is conserved, and a compressional wave can propagate in the skeleton only if $\sin\vartheta_w<c_l/c_s$, whatever lies between water and skeleton. The critical angle $\arcsin(c_l/c_s)$ equals $30^\circ$ for the reference values. Beyond it the compressional field in the skeleton is evanescent and loading proceeds through shear and flexural coupling, which an acoustic skeleton cannot represent. Every range in this study is therefore a vertical standoff.

Consider first a canopy of one-way travel time $\tau\equiv h/U$ lying on a skeletal half-space. The incident pulse enters the canopy with factor $T_1\equiv T_{wc}$, crosses it in time $\tau$, and enters the skeleton with factor $T_2\equiv T_{cs}$. The remainder reflects with $R_{cs}$, returns to the top of the canopy, reflects with $R_{cw}$, and arrives again at the skeleton after a further delay $2\tau$. Each round trip multiplies the amplitude by $q\equiv R_{cs}R_{cw}$, and summing all paths gives \cite{Claerbout1968,Goupillaud1961}
\begin{equation}
p_s(t)=T_1T_2\sum_{n=0}^{\infty}q^n\,p_i(t-\tau-2n\tau).
\label{eq:rayseries}
\end{equation}
Because $Z_c$ is lower than both $Z_w$ and $Z_s$, both reflection coefficients are positive and $0<q<1$. For the exponential pulse \eqref{eq:similitude}, the transmitted pressure just after the arrival of the $N$-th reverberation, at $t=\tau+2N\tau$, is a finite geometric sum,
\begin{equation}
p_s=T_1T_2P\sum_{n=0}^{N}q^n r^{N-n}=T_1T_2P\,a_N,\qquad a_N\equiv\frac{r^{N+1}-q^{N+1}}{r-q},\qquad r\equiv e^{-2\tau/\theta}.
\label{eq:aN}
\end{equation}
Between arrivals $p_s$ decays, so the transmitted peak is $T_1T_2P\max_Na_N$. As $\tau/\theta\to0$, $r\to1$ and $a_N\to(1-q^{N+1})/(1-q)$, whose supremum $1/(1-q)$ recovers the gas-free value. As $\tau/\theta\to\infty$, $r\to0$ and the peak tends to $T_1T_2$.

The transmitted impulse obeys an exact invariant. Integrating Eq.~\eqref{eq:rayseries} term by term gives $\int p_s\,dt=T_1T_2(1-q)^{-1}\int p_i\,dt$. Writing $\zeta_w=Z_w$, $\zeta_c=Z_c$, $\zeta_s=Z_s$,
\begin{equation}
1-q=\frac{(\zeta_s+\zeta_c)(\zeta_w+\zeta_c)-(\zeta_s-\zeta_c)(\zeta_w-\zeta_c)}{(\zeta_s+\zeta_c)(\zeta_w+\zeta_c)}=\frac{2\zeta_c(\zeta_s+\zeta_w)}{(\zeta_s+\zeta_c)(\zeta_w+\zeta_c)},
\label{eq:oneminusq}
\end{equation}
so that $T_1T_2/(1-q)=2Z_s/(Z_s+Z_w)$, independent of $Z_c$ and $\tau$. The result extends to any stack. In the frequency domain, with time dependence $e^{\mathrm{i}\omega t}$, the solution \eqref{eq:dalembert} in a layer of thickness $h_j$ maps the state vector $(\hat p,\hat u)^{\mathsf T}$ at its top to that at its bottom through the propagator
\begin{equation}
\mathsf{L}_j(\omega)=\begin{pmatrix}\cos k_jh_j & -\mathrm{i}Z_j\sin k_jh_j\\ -\mathrm{i}Z_j^{-1}\sin k_jh_j & \cos k_jh_j\end{pmatrix},\qquad k_j\equiv\frac{\omega}{c_j},
\label{eq:propagator}
\end{equation}
and the stack is described by the ordered product $\mathsf{M}=\prod_j\mathsf{L}_j$ \cite{Thomson1950,Haskell1953}. In the water above, $\hat p=\hat A(1+\hat{\mathcal R})$ and $\hat u=\hat A(1-\hat{\mathcal R})/Z_w$ for incident spectrum $\hat A$. In the half-space below, of impedance $Z_b$, $\hat p=\hat{\mathcal T}\hat A$ and $\hat u=\hat{\mathcal T}\hat A/Z_b$. As $\omega\to0$ every layer becomes acoustically thin, $\mathsf{M}\to\mathsf{I}$, and the boundary conditions reduce to Eq.~\eqref{eq:rt} with $Z_2=Z_b$. Since $\int p\,dt=\hat p(0)$,
\begin{equation}
\int_{-\infty}^{\infty}p_s\,dt=\frac{2Z_b}{Z_b+Z_w}\int_{-\infty}^{\infty}p_i\,dt
\label{eq:invariant}
\end{equation}
for any finite stack of lossless linear layers. Physically, the zero-frequency content of a pulse has a wavelength far larger than any layer and therefore cannot see the canopy.

Tension arises at the back of the plate. Let the plate carry a downward step-exponential pulse of peak $P_s$ and let the medium below have impedance $Z_c<Z_s$, so that $R_b=(Z_c-Z_s)/(Z_c+Z_s)<0$. Measure $\xi$ upward from the back face. The incident front passes $\xi$, reaches the back face after a time $\xi/c_s$, and the reflected front returns to $\xi$ after a further $\xi/c_s$. At that instant the incident contribution has decayed by $e^{-2\xi/(c_s\theta)}$ and the reflected contribution equals $R_bP_s$, so
\begin{equation}
p(\xi)=P_s\left[R_b+e^{-2\xi/(c_s\theta)}\right].
\label{eq:firstreflection}
\end{equation}
This is the Hopkinson construction for spallation \cite{Grady1988,Antoun2003}, valid until the reflection from the upper face of the plate returns. For $t$ later than the reflected front, both contributions decay with the same factor, so Eq.~\eqref{eq:firstreflection} is the extremum at each $\xi$. It is most negative at $\xi=d$, and tension exists if and only if $|R_b|>e^{-2\delta}$ with $\delta\equiv d/(c_s\theta)$. With $\zeta\equiv Z_c/Z_s$, $|R_b|=(1-\zeta)/(1+\zeta)$, and the inequality $(1-\zeta)/(1+\zeta)>e^{-2\delta}$ rearranges to
\begin{equation}
\frac{Z_c}{Z_s}<\frac{1-e^{-2\delta}}{1+e^{-2\delta}}=\tanh\delta,\qquad d_c\equiv c_s\theta\,\operatorname{artanh}\frac{Z_c}{Z_s}.
\label{eq:onset}
\end{equation}
Plates thinner than $d_c$ carry no first-reflection tension because the reflected rarefaction meets an incident pulse that has not yet decayed enough. If the tensile strength $\sigma_t$ is reached, setting $-p(\xi)=\sigma_t$ in Eq.~\eqref{eq:firstreflection} gives the depth of first failure,
\begin{equation}
\xi^*=\frac{c_s\theta}{2}\ln\frac{1}{|R_b|-\sigma_t/P_s},
\label{eq:scab}
\end{equation}
which sets the thickness of the spall scab, and spall requires $\xi^*<d$.

The relaxed closure assumes that bubbles follow the local pressure instantaneously. Its validity is assessed from the radial dynamics of a single bubble. For an incompressible liquid, mass conservation gives the radial velocity $u_r=R^2\dot R/r^2$ outside a bubble of radius $R(t)$. Substituting into the radial momentum equation $\partial_tu_r+u_r\partial_ru_r=-\rho_l^{-1}\partial_rp$ and integrating from $r=R$ to infinity yields the Rayleigh-Plesset equation $R\ddot R+\tfrac32\dot R^2=(p_B-p_\infty)/\rho_l$ \cite{PlessetProsperetti1977,Brennen2013}. Retaining liquid compressibility to first order in the wall Mach number $\dot R/c_l$ adds acoustic radiation and gives the Keller-Miksis equation \cite{KellerMiksis1980,LauterbornKurz2010},
\begin{equation}
\left(1-\frac{\dot R}{c_l}\right)R\ddot R+\frac{3}{2}\left(1-\frac{\dot R}{3c_l}\right)\dot R^2=\left(1+\frac{\dot R}{c_l}\right)\frac{p_B-p_\infty(t)}{\rho_l}+\frac{R}{\rho_lc_l}\frac{dp_B}{dt}.
\label{eq:km}
\end{equation}
The liquid pressure at the wall follows from the normal stress balance across the interface, with polytropic gas pressure, Laplace pressure, and the viscous normal stress $2\mu\,\partial_ru_r=-4\mu\dot R/R$,
\begin{equation}
p_B=\left(p_0+\frac{2\sigma}{R_0}\right)\left(\frac{R_0}{R}\right)^{3\kappa}-\frac{2\sigma}{R}-\frac{4\mu\dot R}{R},\qquad p_\infty(t)=p_0+p_i(t),
\label{eq:pb}
\end{equation}
where $\sigma$ is surface tension, $\mu$ liquid viscosity, and $R_0$ the equilibrium radius \cite{LauterbornKurz2010}. Scaling lengths by $R_0$, pressures by $p_0$, and time by $t_c\equiv R_0\sqrt{\rho_l/p_0}$ introduces the Mach number $\mathrm{M}\equiv\sqrt{p_0/\rho_l}/c_l$, the Weber number $\mathrm{We}\equiv2\sigma/(R_0p_0)$, and the Reynolds number $\mathrm{Re}\equiv R_0\sqrt{\rho_lp_0}/\mu$. Setting $R=R_0(1+x)$ with $|x|\ll1$, $\mathrm{M}\to0$, and $\mathrm{Re}\to\infty$, and expanding $p_B$ to first order gives $\ddot x+\omega_0^2x=0$ with
\begin{equation}
\omega_0^2=3\kappa(1+\mathrm{We})-\mathrm{We},
\label{eq:minnaert}
\end{equation}
the Minnaert frequency corrected for surface tension, in units of $t_c^{-1}$ \cite{Minnaert1933}. The product $\theta f_M$, with $f_M\equiv\omega_0/(2\pi t_c)$, compares the pulse duration with the bubble period. The relaxed closure requires $\theta f_M\gg1$ and no inertial overshoot.

Finally, the pattern of loading over an irregular reef is examined with two-dimensional linear acoustics in a vertical plane. With spatially variable density $\rho(\mathbf{x})$ and bulk modulus $K(\mathbf{x})\equiv\rho c^2$, linearizing Eqs.~\eqref{eq:euler} in two dimensions and adding a volume source $q(t)g(\mathbf{x})$ per unit length gives
\begin{equation}
\rho\,\partial_t\mathbf{u}=-\nabla p,\qquad \partial_tp=-K\,\nabla\cdot\mathbf{u}+K\,q(t)\,g(\mathbf{x}).
\label{eq:acoustic2d}
\end{equation}
In a homogeneous region the velocity potential $\phi$, with $\mathbf{u}=\nabla\phi$ and $p=-\rho\,\partial_t\phi$, satisfies $\nabla^2\phi-c^{-2}\partial_t^2\phi=q(t)\delta(\mathbf{x})$ for a point source in the plane. Its solution is the convolution of $q$ with the two-dimensional Green function \cite{Pierce2019},
\begin{equation}
\phi(r,t)=-\frac{1}{2\pi}\int_{-\infty}^{t-r/c}\frac{q(t')\,dt'}{\sqrt{(t-t')^2-r^2/c^2}} .
\label{eq:green2d}
\end{equation}
Far from the source, where $t-r/c\ll r/c$, the square root is approximated by $\sqrt{2r/c}\,\sqrt{t-r/c-t'}$, and the integral becomes proportional to the Riemann-Liouville half-integral $I^{1/2}q$ evaluated at retarded time. Differentiating in time to obtain $p$ then gives
\begin{equation}
p(r,t)\simeq\frac{\rho}{2}\sqrt{\frac{c}{2\pi r}}\;D^{1/2}q\!\left(t-\frac{r}{c}\right),\qquad I^{1/2}f(t)\equiv\frac{1}{\sqrt{\pi}}\int_0^t\frac{f(t')}{\sqrt{t-t'}}\,dt',\quad D^{1/2}\equiv\frac{d}{dt}I^{1/2},
\label{eq:farfield}
\end{equation}
where $D^{1/2}$ is the half-order derivative \cite{Diethelm2010}. A line source therefore radiates the half-derivative of its volume rate, in contrast with the first derivative radiated by a point source in three dimensions. Choosing $q=I^{1/2}[p_i]$ makes $D^{1/2}q=p_i$ and reproduces the target pulse shape in the far field up to a constant factor. The near field retains a slowly decaying component that varies logarithmically with distance, which is a property of line sources and not of the charge. In these runs the canopy is assigned the secant impedance at 5~MPa, the skeleton is again acoustic, and only patterns and ratios that do not depend on source amplitude are interpreted.

\subsection{Numerical Implementation}

The layered column is advanced with the Goupillaud scheme \cite{Goupillaud1961}, which discretizes the characteristic form of Eqs.~\eqref{eq:acoustic1d}. In each layer the Riemann invariants $p\pm Z_ju$ are constant along the characteristics $dz/dt=\pm c_j$, and they equal twice the down-going and up-going amplitudes $f$ and $g$ of Eq.~\eqref{eq:dalembert}. Dividing every layer into cells of length $c_j\Delta t$ makes the one-way travel time across every cell equal to the time step, so each amplitude moves exactly one cell per step and is scattered only at cell boundaries. Denoting by $d_j^n$ and $u_j^n$ the down-going and up-going pressure amplitudes in cell $j$ at step $n$, the interface conditions \eqref{eq:rt} give the scattering relations
\begin{equation}
\begin{pmatrix}d_{j+1}^{\,n+1}\\ u_j^{\,n+1}\end{pmatrix}=\begin{pmatrix}1+R_j & -R_j\\ R_j & 1-R_j\end{pmatrix}\begin{pmatrix}d_j^{\,n}\\ u_{j+1}^{\,n}\end{pmatrix},\qquad R_j\equiv\frac{Z_{j+1}-Z_j}{Z_{j+1}+Z_j},
\label{eq:goupillaud}
\end{equation}
where the up-going wave from below sees the coefficient $-R_j$ and transmits with $1-R_j$. The cell pressure is $p_j^n=d_j^n+u_j^n$. The incident pulse is injected as $d_0^n=p_i(n\Delta t)$, and waves leaving the top of the first cell or the bottom of the last cell are absorbed. Because every delay in the column is an integer multiple of $\Delta t$, the scheme has no dispersion and reproduces the continuous solution exactly at the sampled times, provided layer thicknesses are integer numbers of cells. Thicknesses are rounded to whole cells with $\Delta t$ equal to one hundredth of the smallest decay constant in a sweep, which realizes the reference 12~cm plate as 12.06~cm and the 15~cm canopy to within 0.6~mm. The scheme is batched so that every column in a parameter sweep advances in one time loop, with the impedance of each cell stored per column.

Two further solvers share no code with the first. The ray series \eqref{eq:rayseries} is summed directly on the same time grid until $q^n$ falls below $10^{-18}$. The transfer-matrix solution evaluates the product of propagators \eqref{eq:propagator} at each angular frequency, imposes the radiation conditions stated above Eq.~\eqref{eq:invariant}, and solves the resulting linear system
\begin{equation}
\begin{pmatrix}M_{11}-M_{12}/Z_w & -1\\ M_{21}-M_{22}/Z_w & -Z_b^{-1}\end{pmatrix}\begin{pmatrix}\hat{\mathcal R}\\ \hat{\mathcal T}\end{pmatrix}=-\begin{pmatrix}M_{11}+M_{12}/Z_w\\ M_{21}+M_{22}/Z_w\end{pmatrix}
\label{eq:tmsystem}
\end{equation}
for the transmission coefficient $\hat{\mathcal T}(\omega)$. The transmitted history is recovered by inverse fast Fourier transform of $\hat{\mathcal T}\hat p_i$ with eightfold zero padding, which suppresses wrap-around of the reverberation tail.

The Keller-Miksis equation is written as a first-order system for $(R,\dot R)$. Because $dp_B/dt$ contains $\ddot R$ through the viscous term, all terms in $\ddot R$ are collected on the left, which in dimensionless form gives the coefficient $(1-\mathrm{M}\dot R)R+4\mathrm{M}/\mathrm{Re}$ multiplying $\ddot R$. The system is integrated with the eighth-order Dormand-Prince method at relative tolerance $10^{-10}$ \cite{DormandPrince1980,Hairer1993}. The minimum radius is located by an event on $\dot R=0$ with $\ddot R>0$ rather than read from output samples, which matters because collapse minima are sharp. Results are checked against the implicit Radau method \cite{Hairer1996}.

Equations \eqref{eq:acoustic2d} are discretized on a staggered grid with spacing $\Delta x$ in both directions, pressure at cell centers, and velocity components on cell faces \cite{Yee1966,Virieux1986}. With buoyancy $b\equiv1/\rho$ averaged arithmetically onto faces, one step reads
\begin{align}
u_{i+1/2,k}^{\,n+1/2}&=u_{i+1/2,k}^{\,n-1/2}-\frac{\Delta t\,b_{i+1/2,k}}{\Delta x}\left(p_{i+1,k}^{\,n}-p_{i,k}^{\,n}\right),\label{eq:fdtdu}\\
p_{i,k}^{\,n+1}&=p_{i,k}^{\,n}-\frac{\Delta t\,K_{i,k}}{\Delta x}\left(u_{i+1/2,k}^{\,n+1/2}-u_{i-1/2,k}^{\,n+1/2}+w_{i,k+1/2}^{\,n+1/2}-w_{i,k-1/2}^{\,n+1/2}\right)+\Delta t\,K_{i,k}\,q^{\,n+1/2}g_{i,k},\label{eq:fdtdp}
\end{align}
with the analogous update for the vertical velocity $w$. A von Neumann analysis of the homogeneous scheme gives the stability condition $c\,\Delta t/\Delta x\le1/\sqrt2$, and we use $\Delta t=\Delta x/(2c_{\max})$. The scheme is second order for smooth coefficients and first order across material interfaces. Sponge layers forty cells wide multiply all fields by $\exp[-(s\,m)^2]$ at each step, where $m$ is the distance in cells from the inner edge of the sponge and $s=0.015$ \cite{Cerjan1985}. The upper boundary is absorbing rather than pressure-release because the surface-reflected rarefaction is capped near $-p_0$ by bulk cavitation, which a linear solver cannot represent \cite{Cole1948,Saila1993}. The grid spacing is 4~mm over a domain of $3.2\times2.3$~m, and the reef contains a branching thicket of 3~cm fingers at roughly 9~cm spacing and a 10~cm tabular plate on a stalk, embedded in a 0.36~m canopy over a skeletal base. The half-integral in Eq.~\eqref{eq:farfield} is evaluated by midpoint product integration, which is exact for piecewise-constant integrands.

Damage is assessed from the largest compression and tension reached in the plate within an integration window extending ten decay constants beyond first arrival. A plate is classed as crushed when compression exceeds $\sigma_c$ and as spalled when tension exceeds $\sigma_t$. The spall and crush standoffs are the largest standoffs at which these thresholds are reached, located by linear interpolation in $\log R$. The canopy void fraction is prescribed over the day as
\begin{equation}
\alpha(t)=\alpha_n+(\alpha_{\max}-\alpha_n)\,s(t)^2,\qquad s(t)=\max\left\{0,\sin\frac{\pi(t-6)}{12}\right\}\ \text{for}\ 6<t<18,\quad s(t)=0\ \text{otherwise},
\label{eq:diel}
\end{equation}
with $t$ in local solar hours and $\alpha_n=10^{-5}$. The squared sine delays the appearance of free gas until dissolved oxygen has passed saturation. The peak $\alpha_{\max}$ is a scenario parameter because no void-fraction measurement in a coral canopy is known to us. All computations use NumPy, SciPy, and Matplotlib \cite{Harris2020,Virtanen2020,Hunter2007}. Reference values are listed in Table~\ref{tab:params}.

\begin{table}[H]
\centering
\caption{Reference parameters. Material constants are illustrative and are not calibrated against measurements on a specific reef.}
\label{tab:params}
\begin{tabular}{llr}
\hline
Quantity & Symbol & Value \\
\hline
Seawater density, sound speed & $\rho_l$, $c_l$ & 1025 kg\,m$^{-3}$, 1500 m\,s$^{-1}$ \\
Ambient pressure at 5 m & $p_0$ & 151.6 kPa \\
Gas polytropic exponent & $\kappa$ & 1.4 \\
Skeleton density, longitudinal speed & $\rho_s$, $c_s$ & 1600 kg\,m$^{-3}$, 3000 m\,s$^{-1}$ \\
Skeleton tensile, compressive strength & $\sigma_t$, $\sigma_c$ & 2 MPa, 20 MPa \\
Charge mass (TNT equivalent) & $W$ & 1 kg \\
Canopy thickness above plate & $h$ & 0.15 m \\
Reference plate thickness & $d$ & 0.12 m \\
Bubble radius, surface tension, viscosity & $R_0$, $\sigma$, $\mu$ & 0.5 mm, 0.072 N\,m$^{-1}$, $10^{-3}$ Pa\,s \\
\hline
\end{tabular}
\end{table}

\subsection{Numerical Experiments and Verification}

The experiments proceed from the canopy to the reef. The Rayleigh-line shock speed \eqref{eq:rayleigh} and crossover range, defined by $P(R^*)=p^*$ and hence $R^*=W^{1/3}(K_P/p^*)^{1/A_P}$, are evaluated for void fractions from $10^{-5}$ to $3\times10^{-2}$. Transmission through a canopy over a skeletal half-space is computed for $\tau/\theta$ from zero to twenty. The onset criterion \eqref{eq:onset} is mapped in the plane of $\delta$ and $Z_c/Z_s$, and $d_c$ is evaluated at three standoffs. Keller-Miksis trajectories are computed for $\theta f_M$ between 0.05 and 20 at overpressures of 3, 10, and 30 times $p_0$. Plate stresses are computed on a grid of 64 standoffs between 0.7 and 20~m and 56 void fractions for plates of 6, 9, 12, and 15~cm. The diel scenario uses $\alpha_{\max}=10^{-3}$, $10^{-2}$, and $3\times10^{-2}$. Two-dimensional runs are carried out for $\alpha=10^{-5}$ and $10^{-2}$.

Verification uses independent algorithms and analytical limits. For the layered solvers, a smooth two-exponential pulse $p_i\propto e^{-t/\theta}-e^{-10t/\theta}$ is used so that the Fourier solution is not affected by the discontinuous front, and agreement is measured by the maximum absolute difference $\max_n|p_s^{(a)}(t_n)-p_s^{(b)}(t_n)|$ for a pulse of unit peak. The impulse invariant \eqref{eq:invariant}, the peak formula \eqref{eq:aN}, and the first-reflection profile \eqref{eq:firstreflection} are checked against the Goupillaud solution. Since the Taylor expansion of Eq.~\eqref{eq:dv} predicts $|U/c_W-1|=O(\Delta p)$, the Wood limit is checked by the least-squares slope of $\log|U/c_W-1|$ against $\log\Delta p$ over $10^{-4}\le\Delta p/p_0\le10^{-2.5}$, whose expected value is one.

For the Keller-Miksis solver, a pressure step of relative size $\epsilon$ is applied in the limit $\mathrm{M}\to0$, $\mathrm{Re}\to\infty$. The bubble then oscillates about the displaced equilibrium $R_\epsilon$, whose linear frequency follows from $\omega_\epsilon^2=-R_\epsilon^{-1}\,dp_B/dR|_{R_\epsilon}$. The oscillation amplitude is proportional to $\epsilon$, and a Lindstedt-Poincar\'{e} expansion of a conservative oscillator with quadratic and cubic nonlinearity shows that the frequency correction is quadratic in amplitude \cite{NayfehMook1979}. The measured period, obtained from successive event times of minimum radius, should therefore give $|\omega/\omega_\epsilon-1|\propto\epsilon^2$. For the two-dimensional solver, a quasi-one-dimensional column is run at grid spacings of 4, 2, 1, and 0.5~mm with a smooth Gaussian source, and the observed order is computed from successive differences as $\log_2(e_{h}/e_{h/2})$, with $e_h$ the maximum difference between the solutions at spacings $h$ and $h/2$. The impulse invariant is measured in the same column by comparing the transmitted impulse with that of a run without canopy or skeleton, with the domain long enough that no sponge reflection returns within the window.

\section{Results}

\begin{figure}[H]
\centering
\includegraphics[width=0.9\textwidth]{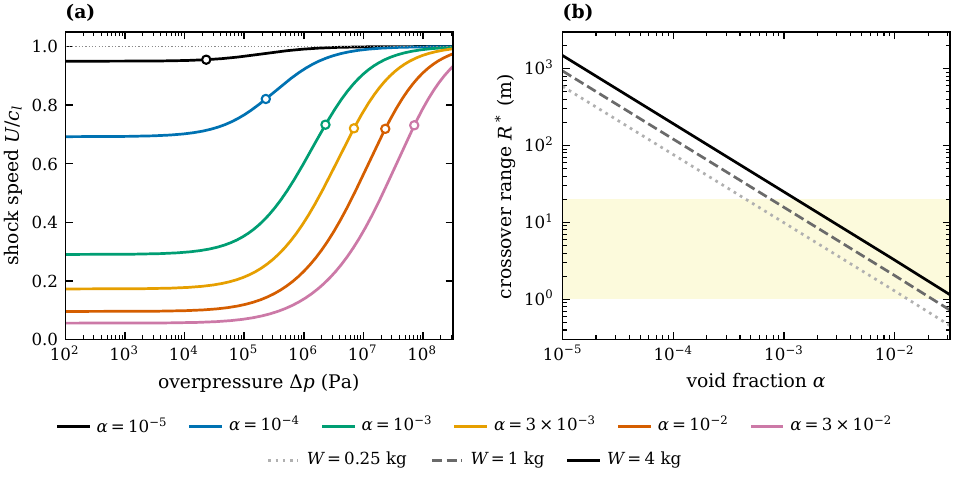}
\caption{The canopy under shock loading. (a) Rayleigh-line shock speed over the liquid sound speed against overpressure. Curves run black, blue, green, orange, vermilion, and purple for $\alpha=10^{-5}$, $10^{-4}$, $10^{-3}$, $3\times10^{-3}$, $10^{-2}$, and $3\times10^{-2}$; the open circle on each curve marks the crossover overpressure $p^*$, and the horizontal grey dotted line is $U=c_l$. (b) Crossover range $R^*$ against void fraction for charges of 0.25, 1, and 4~kg TNT equivalent, as a light grey dotted, mid grey dashed, and black solid line. The pale yellow band spans standoffs of 1 to 20~m.}
\label{fig:canopy}
\end{figure}

The Rayleigh-line shock speed of the canopy rises from the Wood speed at small overpressure toward the liquid sound speed at large overpressure (Fig.~\ref{fig:canopy}a). At 5~m depth, where $p_0=151.6$~kPa, the Wood speed is 0.950, 0.692, 0.290, 0.173, 0.096, and 0.056 times $c_l$ for $\alpha=10^{-5}$, $10^{-4}$, $10^{-3}$, $3\times10^{-3}$, $10^{-2}$, and $3\times10^{-2}$. The corresponding crossover overpressures are 0.023, 0.23, 2.31, 6.94, 23.3, and 71.3~MPa. For $\alpha\geq10^{-3}$ the shock speed at $p^*$ lies between 0.719 and 0.733 times $c_l$, in agreement with Eq.~\eqref{eq:ustar}. The crossover range $R^*$ at which the similitude peak equals $p^*$ falls from 930~m at $\alpha=10^{-5}$ to 15.8~m at $\alpha=10^{-3}$ and 2.04~m at $\alpha=10^{-2}$ for a 1~kg charge (Fig.~\ref{fig:canopy}b). Ranges of a few meters therefore place typical canopy void fractions on either side of the crossover, which means the canopy can be close to transparent for one charge and strongly compliant for another. The largest values of $R^*$ lie far outside the range over which the similitude constants were calibrated and are shown only to display the scaling.

\begin{figure}[H]
\centering
\includegraphics[width=\textwidth]{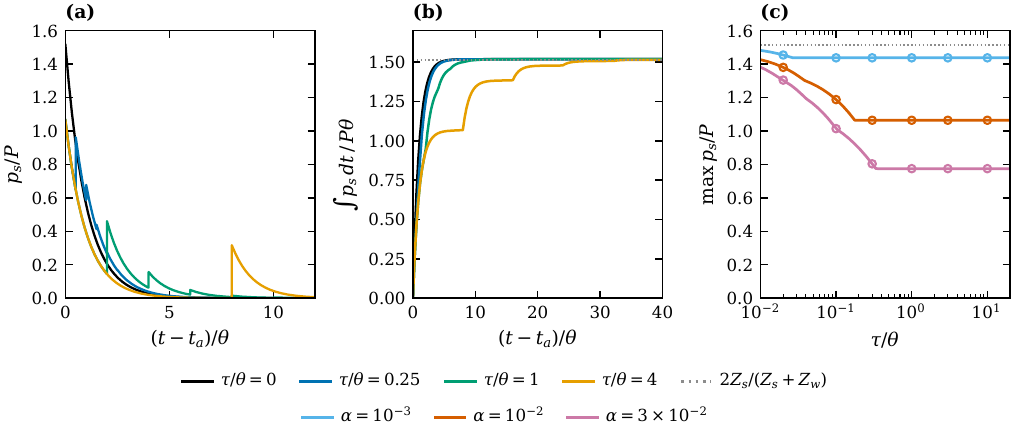}
\caption{Transmission into a skeletal half-space below a canopy with $\alpha=10^{-2}$ and secant impedance at 5~MPa. (a) Transmitted pressure over the incident peak against time since first arrival, with $\tau/\theta=0$, 0.25, 1, and 4 in black, blue, green, and orange. (b) Cumulative transmitted impulse over incident impulse in the same four colours; the horizontal grey dotted line is $2Z_s/(Z_s+Z_w)$. (c) Peak transmitted pressure against $\tau/\theta$ for $\alpha=10^{-3}$, $10^{-2}$, and $3\times10^{-2}$ in light blue, vermilion, and purple. Solid lines are Eq.~\eqref{eq:aN} and open circles are the Goupillaud solution; the grey dotted line is again the impulse invariant.}
\label{fig:transmission}
\end{figure}

Transmission through a canopy into a skeletal half-space follows the structure of Eq.~\eqref{eq:rayseries} (Fig.~\ref{fig:transmission}). Without a canopy the transmitted peak is $2Z_s/(Z_s+Z_w)=1.515$ times the incident peak. As the canopy travel time grows, the pulse splits into a train of reverberations whose ratio at 5~MPa is $q=0.051$, 0.297, and 0.489 for $\alpha=10^{-3}$, $10^{-2}$, and $3\times10^{-2}$. The transmitted peak falls toward $T_1T_2=1.438$, 1.064, and 0.775 for the same void fractions and reaches that limit once $\tau/\theta$ exceeds a few tenths. The cumulative transmitted impulse converges to 1.515 times the incident impulse for every canopy thickness, as Eq.~\eqref{eq:invariant} requires (Fig.~\ref{fig:transmission}b). A gas-laden canopy thus lowers the peak compression delivered to the skeleton by up to about half while leaving the impulse unchanged.

\begin{figure}[H]
\centering
\includegraphics[width=0.9\textwidth]{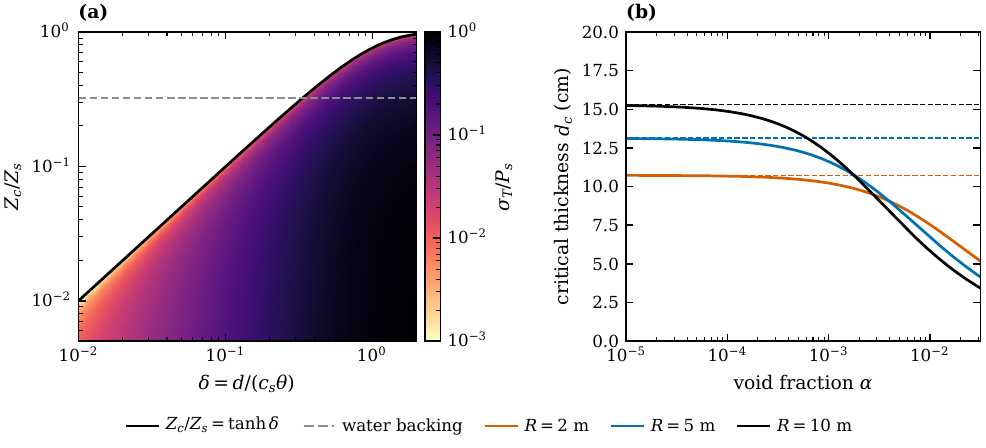}
\caption{Spall onset. (a) First-reflection tensile peak over the transmitted peak, $|R_b|-e^{-2\delta}$, against $\delta\equiv d/(c_s\theta)$ and $Z_c/Z_s$, shaded on a logarithmic scale that runs from pale cream at low values to black at high values. Tension exists only below the solid black curve $Z_c/Z_s=\tanh\delta$; the region above it is left unshaded. The horizontal grey dashed line marks a water-backed plate. (b) Critical thickness $d_c$ against void fraction for a 1~kg charge at standoffs of 2, 5, and 10~m, drawn as solid vermilion, blue, and black curves, with the gas-free value at each standoff as a thin dashed line of the same colour.}
\label{fig:spall}
\end{figure}

The onset criterion \eqref{eq:onset} separates the plane of $\delta$ and $Z_c/Z_s$ into a region without first-reflection tension and a region in which the tensile peak grows to a large fraction of the transmitted peak (Fig.~\ref{fig:spall}a). A water-backed plate has $Z_w/Z_s=0.320$ at the reference values, so it carries tension only when $\delta$ exceeds $\operatorname{artanh}(0.320)=0.332$. For a 1~kg charge, the critical thickness of a water-backed plate is 10.7, 13.1, and 15.3~cm at standoffs of 2, 5, and 10~m, because $\theta$ lengthens with standoff (Fig.~\ref{fig:spall}b). At $\alpha=10^{-2}$ the critical thickness falls to 7.5, 6.7, and 5.8~cm, and at $\alpha=3\times10^{-2}$ to 5.3, 4.2, and 3.5~cm. The curves for different standoffs cross because two effects oppose each other. At short standoff the peak is large, the canopy is stiffened toward transparency, and $Z_c$ rises. At long standoff $\theta$ is longer and the plate is thinner relative to $c_s\theta$. At a standoff of 3~m and $\alpha=10^{-2}$, Eq.~\eqref{eq:scab} places first failure 9.8~cm behind the back face for a plate loaded through water, which indicates that a 12~cm plate would lose most of its thickness as a single scab under these idealized conditions.

\begin{figure}[H]
\centering
\includegraphics[width=0.9\textwidth]{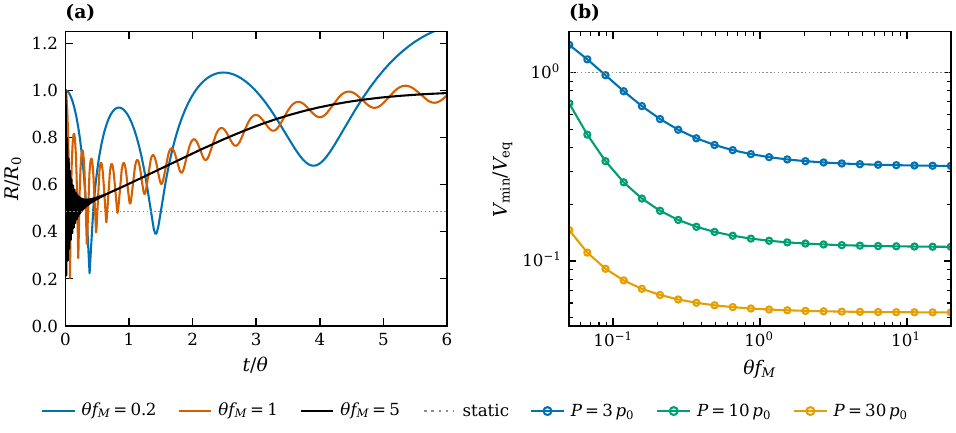}
\caption{Keller-Miksis dynamics of a 0.5~mm canopy bubble. (a) Radius over the equilibrium radius under a step-exponential overpressure of $20p_0$, with $\theta f_M=0.2$, 1, and 5 in blue, vermilion, and black; the horizontal grey dotted line is the static radius at the peak pressure. (b) Minimum volume over the static volume at peak pressure against $\theta f_M$ for overpressures of $3p_0$, $10p_0$, and $30p_0$, drawn as open circles joined by lines in blue, green, and orange. The horizontal grey dotted line marks unity, where the minimum volume would equal the static value.}
\label{fig:bubble}
\end{figure}

The Keller-Miksis calculations test the relaxed closure on which these results rest (Fig.~\ref{fig:bubble}). A bubble of radius 0.5~mm at 5~m depth has a surface-tension-corrected Minnaert frequency of 7.94~kHz, so the similitude decay constants of 0.1 to 0.15~ms give $\theta f_M$ of order one. Under a step overpressure the bubble overshoots its static radius and rings (Fig.~\ref{fig:bubble}a). The minimum volume reached is 0.32 to 1.40 times the static volume at the peak pressure for an overpressure of $3p_0$, 0.12 to 0.69 times for $10p_0$, and 0.054 to 0.15 times for $30p_0$ over $0.05\leq\theta f_M\leq20$ (Fig.~\ref{fig:bubble}b). The ratio does not approach one for long pulses, because the abrupt front drives the bubble past equilibrium regardless of pulse length. The canopy is therefore neither frozen nor relaxed on the time scale of the pulse.

\begin{figure}[H]
\centering
\includegraphics[width=\textwidth]{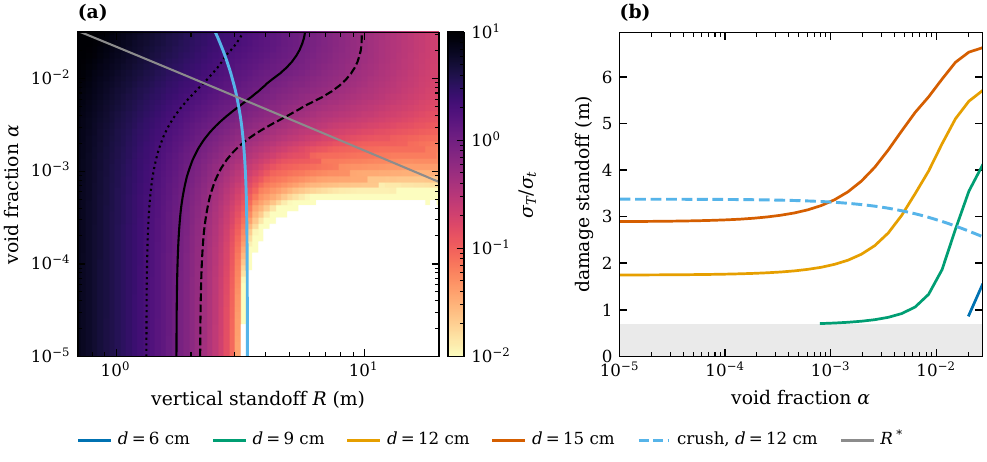}
\caption{Damage regimes for a 1~kg charge directly above the plate. (a) Largest tensile stress within the integration window in a 12~cm plate over its tensile strength, against vertical standoff and void fraction, shaded on a logarithmic scale from pale cream at low values to black at high values, with values below one hundredth left unshaded. Solid, dashed, and dotted black contours mark $\sigma_T=\sigma_t$ for $\sigma_t=2$, 1, and 4~MPa. The light blue contour encloses compressive failure at $\sigma_c=20$~MPa, and the solid grey line is $R^*(\alpha)$. (b) Spall standoff against void fraction for plates of 6, 9, 12, and 15~cm, as solid blue, green, orange, and vermilion curves, with the crush standoff of the 12~cm plate as a light blue dashed curve. The grey shaded band lies below the smallest standoff computed, so a curve reaching its upper edge begins inside it.}
\label{fig:regimes}
\end{figure}

The full column combines shielding by the upper canopy with softening of the lower one (Fig.~\ref{fig:regimes}). For a 12~cm plate under a 1~kg charge directly overhead, the spall standoff increases from 1.75~m at $\alpha=10^{-5}$ to 5.70~m at $\alpha=2.7\times10^{-2}$, while the crush standoff decreases from 3.37~m to 2.51~m. Above $\alpha$ of a few times $10^{-3}$ the spall standoff therefore exceeds the crush standoff, and the plate fails in tension at standoffs where it would survive in compression. Thinner plates respond only at larger void fractions. A 6~cm plate first spalls beyond 0.7~m only as $\alpha$ approaches $3\times10^{-2}$, reaching 1.54~m, whereas a 9~cm plate reaches 4.10~m and a 15~cm plate 6.62~m, up from 2.89~m. These values follow directly from the critical thickness, which falls below the plate thickness only when the lower canopy is soft enough.

\begin{figure}[H]
\centering
\includegraphics[width=0.5\textwidth]{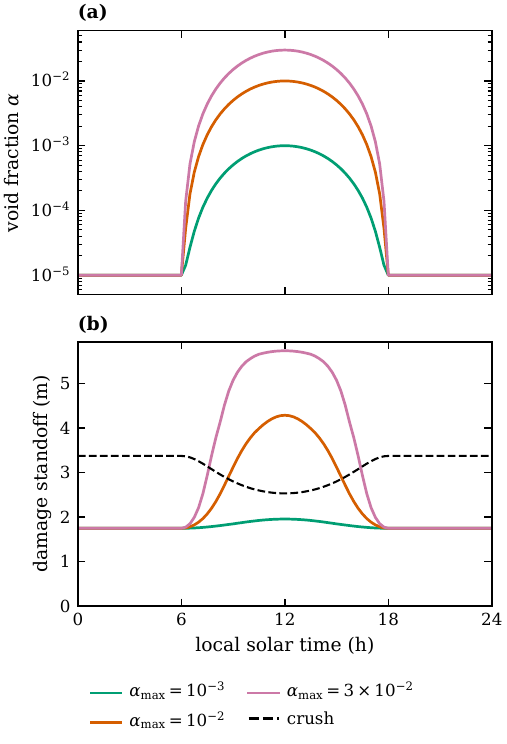}
\caption{Diel cycle for a 1~kg charge directly above plates at 5~m depth. (a) Prescribed canopy void fraction against local solar time for peak values of $10^{-3}$, $10^{-2}$, and $3\times10^{-2}$, in green, vermilion, and purple. (b) Spall standoff of a 12~cm plate in the same three colours, and crush standoff as a black dashed curve for $\alpha_{\max}=3\times10^{-2}$, the scenario in which it varies most.}
\label{fig:diel}
\end{figure}

Under the prescribed diel cycle (Fig.~\ref{fig:diel}), the spall standoff of a 12~cm plate varies between 1.75~m at night and 1.95~m at noon for $\alpha_{\max}=10^{-3}$, a modest change. For $\alpha_{\max}=10^{-2}$ it reaches 4.29~m at noon, and for $\alpha_{\max}=3\times10^{-2}$ it reaches 5.74~m. The crush standoff moves in the opposite direction, from 3.37~m at night to 3.31, 2.93, and 2.53~m at noon in the three scenarios. The same charge detonated at the same height would therefore cause different damage at different times of day if canopy void fractions approach $10^{-2}$, and nearly the same damage if they remain near $10^{-3}$.

\begin{figure}[H]
\centering
\includegraphics[width=\textwidth]{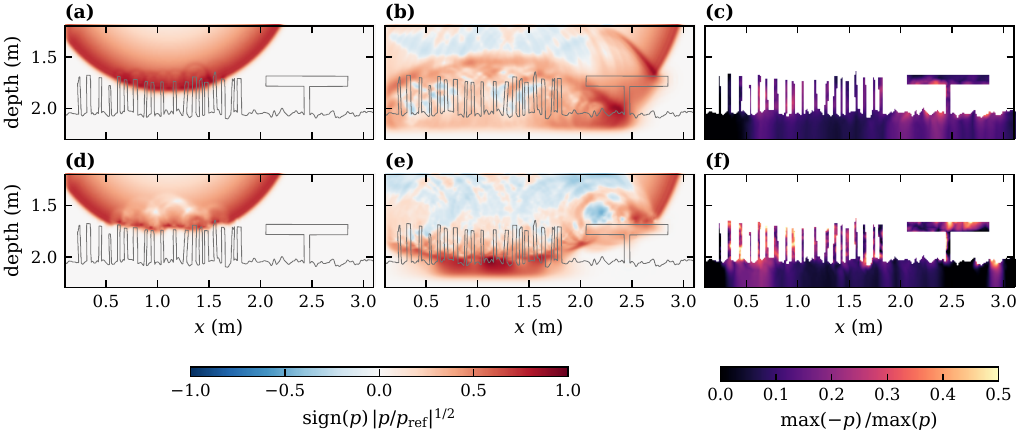}
\caption{Two-dimensional linear acoustics over a branching thicket and a 10~cm tabular plate, with $\alpha=10^{-5}$ (top row) and $\alpha=10^{-2}$ (bottom row). (a, b, d, e) Pressure at two instants shown as $\mathrm{sign}(p)|p/p_{\mathrm{ref}}|^{1/2}$ on a diverging red and blue scale, red for compression and blue for tension, with $p_{\mathrm{ref}}$ the 99.5th percentile of peak compression in the reef of the gas-poor run. The thin grey line is the outline of the skeleton. Blue patches above the reef belong to the late-time near field of the line source and are identical in both runs. (c, f) Ratio of peak tension to peak compression within the skeleton, shaded from black at zero to pale yellow at one half, with water left blank.}
\label{fig:fdtd}
\end{figure}

The two-dimensional runs reproduce the same contrast over an irregular reef (Fig.~\ref{fig:fdtd}). With a gas-rich canopy, the incident front is delayed and weakened inside the canopy, and reverberations fill the canopy with alternating compression and tension. Over the emergent skeleton, the median peak compression relative to the gas-poor reference falls from 0.555 to 0.329 when $\alpha$ rises from $10^{-5}$ to $10^{-2}$. The median ratio of peak tension to peak compression rises from 0.137 to 0.193, the 90th percentile from 0.216 to 0.348, and the 99th percentile from 0.314 to 0.512. These ratios do not depend on source amplitude. Tension concentrates in the tabular plate and in the fingers of the thicket, where the skeleton is surrounded by canopy on more than one side.

\begin{figure}[H]
\centering
\includegraphics[width=0.85\textwidth]{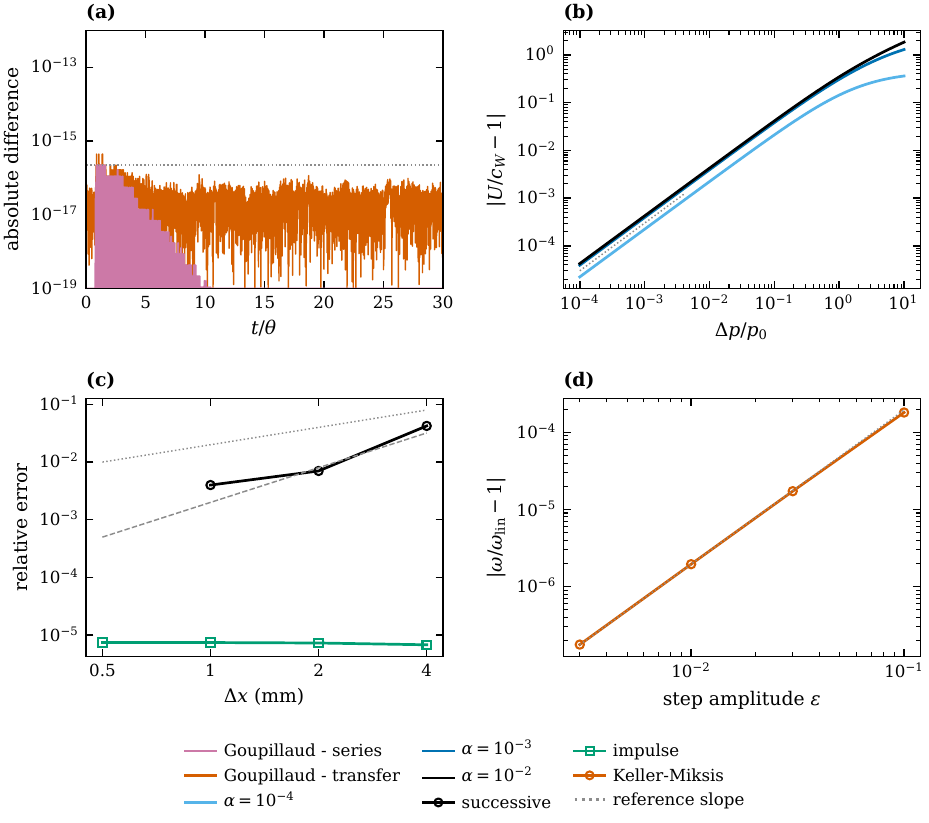}
\caption{Verification. (a) Absolute difference between the Goupillaud solution and the ray series in purple, and between the Goupillaud and transfer-matrix solutions in vermilion; the horizontal grey dotted line is machine epsilon. (b) Relative departure of the Rayleigh-line speed from the Wood speed against overpressure for $\alpha=10^{-4}$, $10^{-3}$, and $10^{-2}$ in light blue, blue, and black, with a grey dotted reference line of slope one. (c) Self-convergence of the two-dimensional solver as black open circles and the error of the impulse invariant as green open squares, with grey dotted and grey dashed reference lines of slope one and two. (d) Relative departure of the Keller-Miksis frequency from the linear frequency about the displaced equilibrium, as vermilion open circles, with a grey dotted reference line of slope two.}
\label{fig:verification}
\end{figure}

The verification results are summarized in Fig.~\ref{fig:verification}. The Goupillaud solution differs from the ray series by at most $2.2\times10^{-16}$ and from the transfer-matrix solution by at most $4.4\times10^{-16}$ for a pulse of unit peak. All three reproduce the impulse invariant to a relative error of $2.2\times10^{-16}$, the peak formula \eqref{eq:aN} is reproduced exactly, and the first-reflection profile \eqref{eq:firstreflection} to $2.2\times10^{-16}$. The Rayleigh-line speed approaches the Wood speed with slope 0.9998 in overpressure and a relative departure of $3.9\times10^{-5}$ at $\Delta p=10^{-4}p_0$. The nonlinear frequency shift of Keller-Miksis oscillations scales with slope 1.978 in step amplitude, against the expected value of two, and the Dormand-Prince and Radau integrations agree to $4.9\times10^{-13}$ in period and $1.4\times10^{-12}$ in minimum radius. The two-dimensional solver shows successive differences of $4.24\times10^{-2}$, $7.02\times10^{-3}$, and $4.00\times10^{-3}$ relative to the peak signal, with observed orders of 2.59 and 0.81. The first value is pre-asymptotic, and the second approaches one as expected for arithmetic averaging of buoyancy across discontinuous interfaces. Its impulse error remains between $6.8\times10^{-6}$ and $7.5\times10^{-6}$ at every resolution because it is set by truncation of the reverberation tail at the end of the window.

\section{Discussion}

The central result is that a gas-laden canopy changes the type of loading on reef skeleton more than it changes its magnitude. The impulse delivered to a skeletal half-space is fixed by Eq.~\eqref{eq:invariant} and cannot be reduced by any lossless canopy, while the peak compression can be lowered substantially. The same low impedance that shields the upper face of a plate softens its lower face, and once $Z_c/Z_s$ falls below $\tanh\delta$ a transmitted compressive pulse returns as tension. Because brittle porous solids are generally much weaker in tension than in compression \cite{Meyers1994}, and the compressive strength assumed here lies within the range measured for coral skeleton \cite{Chamberlain1978}, this conversion extends the standoff over which a plate fails even as the standoff for compressive failure contracts. The criterion is purely kinematic and holds for any material constants, whereas the standoffs depend on the assumed strengths and should be read as scenarios.

The mechanism is bounded by the crossover overpressure. Close to a charge the canopy is compressed well beyond $p^*$ and becomes nearly transparent, so gas matters least exactly where compressive damage is already certain. Farther away the canopy is compliant, but the incident pulse is weaker. The window in which canopy gas matters is therefore set jointly by $p^*$, the similitude decay of the peak, and the critical thickness, and for a 1~kg charge it spans standoffs of a few meters when void fractions are of order $10^{-2}$. At void fractions near $10^{-3}$ the effect is small for every plate thickness examined. Diel oxygen bubbles are well documented acoustically in seagrass meadows \cite{Felisberto2015}, but we are not aware of void-fraction measurements in coral canopies. The model identifies $\alpha\approx10^{-3}$ as the level that such measurements would need to reach before the mechanism becomes relevant.

The relaxed closure is the most important approximation, and the Keller-Miksis results show that it is not satisfied. With $\theta f_M$ of order one, bubbles cannot follow the pressure within the pulse, and the step front drives them past equilibrium. In a relaxing medium the leading edge of a shock travels near the frozen sound speed and carries little of the jump, while the equilibrium state is reached over a relaxation zone whose length depends on bubble dynamics and relative motion \cite{Noordzij1974,vanWijngaarden1972,Brennen2013}. For the parameters used here that length is comparable to the canopy thickness. Early in the pulse the canopy should therefore present less contrast than its secant impedance implies, which suggests that the reported effects are upper bounds for both front-face shielding and back-face tension. A natural extension replaces the frozen secant by a bracket between the frozen limit, in which the canopy impedance is close to that of water, and the relaxed limit used here.

Several other approximations bias the results in known directions. The canopy below the plate is loaded by the transmitted pulse rather than by the incident peak, so its secant impedance is lower than the value used and the back-face effect is understated. The wave reflected from the top of the canopy is negative, and above the interface the sum of incident tail and reflection becomes tensile within a few centimeters with megapascal amplitude, so the water there should cavitate as it does near a free surface \cite{Cole1948,Saila1993}. Cavitation makes later reverberations lossy and nonlinear, and the impulse invariant then holds only for the linear model. Static tensile strength understates the dynamic spall strength of brittle porous solids at strain rates of order $10^3$~s$^{-1}$ \cite{Grady1988,Antoun2003}, so the spall standoffs are also likely upper bounds. The similitude law for peak pressure is supported by measurements for charges of the relevant size \cite{SolowayDahl2014}, but the decay constant and the later pulse shape are less well constrained, and the ammonium-nitrate charges used in practice differ from TNT in energy release \cite{GeersHunter2002}.

Geometry restricts the results to charges directly above the plate. Beyond the $30^\circ$ critical angle no compressional wave enters an acoustic skeleton, and a real plate is then loaded through shear and flexural coupling. Horizontal damage radii require an elastic treatment of the skeleton at oblique incidence, which the Goupillaud scheme accommodates through vertical impedances and travel times and the propagator solution accommodates through complex vertical wavenumbers \cite{Thomson1950,Haskell1953}. The one-dimensional plate is appropriate for tabular colonies whose lateral extent exceeds $c_s\theta$, roughly 0.3 to 0.5~m here, and not for branching colonies, for which the two-dimensional runs indicate only that tension concentrates where skeleton is surrounded by canopy.

For reef management the implications are qualitative. Empirical damage radii that depend only on charge size omit the state of the reef at the time of the blast, and the model suggests that the density and void fraction of the canopy could modify those radii by a factor of two or more for plates of intermediate thickness. Spall produces fragments of characteristic thickness given by Eq.~\eqref{eq:scab}, and rubble of that size is what becomes mobile and suppresses recovery \cite{Fox2003,FoxCaldwell2006}. Linking fragment size to rubble mobility under ambient waves is a direct next step. Fish mortality, which depends on swim-bladder response and on the structure of schools near the charge, lies outside the present model \cite{Saila1993,Herho2026dewikadita}.

\section{Conclusion}

We formulated an idealized model of blast-fishing shock loading on coral skeleton beneath a gas-laden canopy, derived from the conservation of mass and momentum in a relaxed bubbly mixture and from the linear acoustics of a layered reef column. The model yields closed forms for the shock impedance of the canopy, for the overpressure above which the canopy becomes nearly transparent, for a transmitted impulse that no lossless canopy can change, and for a spall criterion that compares the canopy impedance with a threshold set by plate thickness and pulse duration. Independent solvers agree to machine precision. For a small charge directly overhead, a gas-rich canopy substantially extends the standoff at which a tabular plate fails in tension while shortening the standoff at which it is crushed, and a daily cycle of photosynthetic gas makes the same charge more damaging near noon than at night. The effect is weak when canopy gas is scarce. Bubble dynamics indicate that the canopy does not reach equilibrium within the pulse, so the results are best read as bounds. Measurements of free gas in coral canopies, an elastic treatment of oblique incidence, and a relaxing description of the canopy are the steps needed to turn these bounds into predictions.

\section*{Acknowledgements}

This research was supported by the PPMI Research Program 2026 of the Faculty of Earth Sciences and Technology (FITB), Bandung Institute of Technology (ITB), under Project ID FITB.PPMI-1-19-2026, and by the ITB 3P Research Program (Talenta Unggul Scheme) through the Directorate of Research and Innovation, ITB, under Project ID DRI.PN-6-64-2026.

\section*{Open Research}

The source code, figure scripts, data tables, animations, and plain-text reports underlying this study are available at \url{https://github.com/sandyherho/reefblast} under the MIT license. No external data were used. All results are reproducible by running a single script on a standard desktop computer.


\begin{thebibliography}{99}

\bibitem{Anwar2026balikpapan} Anwar, I. P.; Herho, S. H. S.; Khadami, F.;
Putri, M. R.; Syahrial, S. C. Towards statistical modeling of
chlorophyll-a concentrations in Balikpapan Bay, Indonesia: Implications for
algal bloom detection. \textit{Environ. Res. Commun.} \textbf{2026},
\textit{8(2)}, 025027.
\url{https://doi.org/10.1088/2515-7620/ae4680}

\bibitem{Antoun2003} Antoun, T.; Curran, D. R.; Razorenov, S. V.; Seaman, L.; Kanel, G. I.; Utkin, A. V. \textit{Spall Fracture}; Springer: New York,
NY, USA, 2003.
\url{https://doi.org/10.1007/b97226}

\bibitem{Brennen2013} Brennen, C. E. \textit{Cavitation and Bubble
Dynamics}; Cambridge University Press: Cambridge, UK, 2013.
\url{https://doi.org/10.1017/CBO9781107338760}

\bibitem{CampbellPitcher1958} Campbell, I. J.; Pitcher, A. S. Shock waves
in a liquid containing gas bubbles. \textit{Proc. R. Soc. London Ser. A}
\textbf{1958}, \textit{243(1235)}, 534--545. \url{https://doi.org/10.1098/rspa.1958.0018}

\bibitem{Carstensen1947} Carstensen, E. L.; Foldy, L. L. Propagation of
Sound Through a Liquid Containing Bubbles. \textit{J. Acoust. Soc. Am.}
\textbf{1947}, \textit{19(3)}, 481--501.
\url{https://doi.org/10.1121/1.1916508}

\bibitem{Cerjan1985} Cerjan, C.; Kosloff, D.; Kosloff, R.; Reshef, M. A
nonreflecting boundary condition for discrete acoustic and elastic wave
equations. \textit{Geophysics} \textbf{1985}, \textit{50(4)}, 705--708.
\url{https://doi.org/10.1190/1.1441945}

\bibitem{Chamberlain1978} Chamberlain, J. A. Mechanical properties of coral
skeleton: compressive strength and its adaptive significance.
\textit{Paleobiology} \textbf{1978}, \textit{4(4)}, 419--435.
\url{https://doi.org/10.1017/S0094837300006163}

\bibitem{Claerbout1968} Claerbout, J. F. Synthesis of a layered medium from
its acoustic transmission response. \textit{Geophysics} \textbf{1968},
\textit{33(2)}, 264--269.
\url{https://doi.org/10.1190/1.1439927}

\bibitem{Cole1948} Cole, R. H. \textit{Underwater Explosions}; Princeton
University Press: Princeton, NJ, USA, 1948.

\bibitem{CommanderProsperetti1989} Commander, K. W.; Prosperetti, A. Linear
pressure waves in bubbly liquids: Comparison between theory and experiments.
\textit{J. Acoust. Soc. Am.} \textbf{1989}, \textit{85(2)}, 732--746.
\url{https://doi.org/10.1121/1.397599}

\bibitem{Diethelm2010} Diethelm, K. \textit{The Analysis of Fractional
Differential Equations: An Application-Oriented Exposition Using
Differential Operators of Caputo Type}; Lecture Notes in Mathematics;
Springer: Berlin, Germany, 2010.
\url{https://doi.org/10.1007/978-3-642-14574-2}

\bibitem{DormandPrince1980} Dormand, J. R.; Prince, P. J. A family of
embedded Runge-Kutta formulae. \textit{J. Comput. Appl. Math.}
\textbf{1980}, \textit{6(1)}, 19--26.
\url{https://doi.org/10.1016/0771-050X(80)90013-3}

\bibitem{Edinger1998} Edinger, E. N.; Jompa, J.; Limmon, G. V.;
Widjatmoko, W.; Risk, M. J. Reef degradation and coral biodiversity in
Indonesia: Effects of land-based pollution, destructive fishing practices
and changes over time. \textit{Mar. Pollut. Bull.} \textbf{1998},
\textit{36(8)}, 617--630.
\url{https://doi.org/10.1016/S0025-326X(98)00047-2}

\bibitem{Felisberto2015} Felisberto, P.; Jesus, S. M.; Zabel, F.;
Santos, R.; Silva, J.; Gobert, S.; Beer, S.; Bj\"{o}rk, M.; Mazzuca, S.;
Procaccini, G.; Runcie, J. W.; Champenois, W.; Borges, A. V. Acoustic
monitoring of O$_2$ production of a seagrass meadow. \textit{J. Exp. Mar.
Biol. Ecol.} \textbf{2015}, \textit{464}, 75--87.
\url{https://doi.org/10.1016/j.jembe.2014.12.013}

\bibitem{FoxCaldwell2006} Fox, H. E.; Caldwell, R. L. RECOVERY FROM BLAST FISHING ON CORAL REEFS: A TALE OF TWO SCALES. \textit{Ecol. Appl.}
\textbf{2006}, \textit{16(5)}, 1631--1635.
\url{https://doi.org/10.1890/1051-0761(2006)016[1631:RFBFOC]2.0.CO;2}

\bibitem{Fox2003} Fox, H. E.; Pet, J. S.; Dahuri, R.; Caldwell, R. L.
Recovery in rubble fields: long-term impacts of blast fishing. \textit{Mar.
Pollut. Bull.} \textbf{2003}, \textit{46(8)}, 1024--1031.
\url{https://doi.org/10.1016/S0025-326X(03)00246-7}

\bibitem{GeersHunter2002} Geers, T. L.; Hunter, K. S. An integrated
wave-effects model for an underwater explosion bubble. \textit{J. Acoust.
Soc. Am.} \textbf{2002}, \textit{111(4)}, 1584--1601.
\url{https://doi.org/10.1121/1.1458590}

\bibitem{Goupillaud1961} Goupillaud, P. L. An approach to inverse filtering
of near-surface layer effects from seismic records. \textit{Geophysics}
\textbf{1961}, \textit{26(6)}, 754--760.
\url{https://doi.org/10.1190/1.1438951}

\bibitem{Grady1988} Grady, D. E. The spall strength of condensed matter.
\textit{J. Mech. Phys. Solids} \textbf{1988}, \textit{36(3)}, 353--384.
\url{https://doi.org/10.1016/0022-5096(88)90015-4}

\bibitem{Hairer1993} Hairer, E.; Wanner, G.; N{\o}rsett, S. P.
\textit{Solving Ordinary Differential Equations I: Nonstiff Problems}, 2nd
ed.; Springer: Berlin, Germany, 1993.
\url{https://doi.org/10.1007/978-3-540-78862-1}

\bibitem{Hairer1996} Hairer, E.; Wanner, G. \textit{Solving Ordinary
Differential Equations II: Stiff and Differential-Algebraic Problems}, 2nd
ed.; Springer: Berlin, Germany, 1996.
\url{https://doi.org/10.1007/978-3-642-05221-7}

\bibitem{HamptonSmith2021} Hampton-Smith, M.; Bower, D. S.; Mika, S. A
review of the current global status of blast fishing: Causes, implications
and solutions. \textit{Biol. Conserv.} \textbf{2021}, \textit{262}, 109307.
\url{https://doi.org/10.1016/j.biocon.2021.109307}

\bibitem{Harris2020} Harris, C. R.; Millman, K. J.; van der Walt, S. J.;
Gommers, R.; Virtanen, P.; Cournapeau, D.; Wieser, E.; Taylor, J.;
Berg, S.; Smith, N. J.; Kern, R.; Picus, M.; Hoyer, S.; van Kerkwijk, M. H.;
Brett, M.; Haldane, A.; del R{\'\i}o, J. F.; Wiebe, M.; Peterson, P.;
G{\'e}rard-Marchant, P.; Sheppard, K.; Reddy, T.; Weckesser, W.;
Abbasi, H.; Gohlke, C.; Oliphant, T. E. Array programming with NumPy.
\textit{Nature} \textbf{2020}, \textit{585}, 357--362.
\url{https://doi.org/10.1038/s41586-020-2649-2}

\bibitem{Haskell1953} Haskell, N. A. The dispersion of surface waves on
multilayered media. \textit{Bull. Seismol. Soc. Am.} \textbf{1953},
\textit{43(1)}, 17--34.
\url{https://doi.org/10.1785/BSSA0430010017}

\bibitem{Herho2026dewikadita} Herho, S. H. S.; Anwar, I. P.; Khadami, F.;
Handayani, A. P.; Sujatmiko, K. A.; Kasim, K.; Suwarman, R.; Irawan, D. E.
dewi-kadita: A Python library for idealized fish schooling simulation with
entropy-based diagnostics. \textit{J. Phys. Commun.} \textbf{2026},
\textit{10(6)}, 065002.
\url{https://doi.org/10.1088/2399-6528/ae7177}

\bibitem{Herho2026wave} Herho, S. H. S.; Anwar, I. P.; Khadami, F.; Ndruru,
T. R. E. B. N.; Suwarman, R.; Irawan, D. E. wave-attenuation-1d: An idealized
one-dimensional framework for wave attenuation through coastal vegetation
using Numba-accelerated shallow water equations. \textit{J. Theor. Appl.
Mech.} \textbf{2026}, \textit{56}, 89--102.
\url{https://doi.org/10.55787/jtams.2026.1.AI00236}

\bibitem{Herho2026nlse} Herho, S. H. S.; Anwar, I. P.; Khadami, F.;
Suwarman, R.; Irawan, D. E. simple-idealized-1d-nlse: Pseudo-spectral solver
for the 1D nonlinear Schr\"{o}dinger equation. \textit{Rev. Mex. F\'{i}s. E}
\textbf{2026}, \textit{23(2)}, 020206.
\url{https://doi.org/10.31349/RevMexFisE.23.020206}

\bibitem{Herho2026morowali} Herho, S. H. S.; Handayani, A. P.; Anwar, I. P.;
Khadami, F.; Sujatmiko, K. A.; Wibisono, D. Y.; Suwarman, R.; Irawan, D. E.
Causal attribution of coastal water clarity degradation to nickel
processing expansion at the Indonesia Morowali Industrial Park, Sulawesi.
\textit{Environ. Res. Commun.} \textbf{2026}, \textit{8(6)}, 065058.
\url{https://doi.org/10.1088/2515-7620/ae7b00}

\bibitem{Herho2025kh2d} Herho, S. H. S.; Trilaksono, N. J.; Fajary, F. R.;
Napitupulu, G.; Anwar, I. P.; Khadami, F.; Irawan, D. E. kh2d-solver: A
Python library for idealized two-dimensional incompressible Kelvin-Helmholtz
instability. \textit{Appl. Comput. Mech.} \textbf{2025}, \textit{19(2)},
125--156.
\url{https://doi.org/10.24132/acm.2025.1040}

\bibitem{Hunter2007} Hunter, J. D. Matplotlib: A 2D graphics environment.
\textit{Comput. Sci. Eng.} \textbf{2007}, \textit{9(3)}, 90--95.
\url{https://doi.org/10.1109/MCSE.2007.55}

\bibitem{HunterGeers2004} Hunter, K. S.; Geers, T. L. Pressure and velocity
fields produced by an underwater explosion. \textit{J. Acoust. Soc. Am.}
\textbf{2004}, \textit{115(4)}, 1483--1496.
\url{https://doi.org/10.1121/1.1648680}

\bibitem{Irawan2026amerta} Irawan, D. E.; Herho, S. H. S.; Anwar, I. P.;
Khadami, F.; Pamumpuni, A.; Kartiko, R. D.; Riawan, E.; Suwarman, R.;
Puradimaja, D. J. amerta: A Python library for idealized 1D Saint-Venant
dam-break simulation. \textit{Front. Water} \textbf{2026}, \textit{8},
1900409.
\url{https://doi.org/10.3389/frwa.2026.1900409}

\bibitem{Irawan2026kdv} Irawan, D. E.; Herho, S. H. S.; Pamumpuni, A.;
Kartiko, R. D.; Khadami, F.; Anwar, I. P.; Sujatmiko, K. A.; Handayani,
A. P.; Fajary, F. R.; Suwarman, R. An Open-Source Pseudo-Spectral Solver for Idealized Korteweg–de Vries Soliton Simulations. \textit{Water}
\textbf{2026}, \textit{18(7)}, 779.
\url{https://doi.org/10.3390/w18070779}

\bibitem{KellerMiksis1980} Keller, J. B.; Miksis, M. Bubble oscillations of
large amplitude. \textit{J. Acoust. Soc. Am.} \textbf{1980}, \textit{68(2)},
628--633.
\url{https://doi.org/10.1121/1.384720}

\bibitem{LauterbornKurz2010} Lauterborn, W.; Kurz, T. Physics of bubble
oscillations. \textit{Rep. Prog. Phys.} \textbf{2010}, \textit{73(10)},
106501.
\url{https://doi.org/10.1088/0034-4885/73/10/106501}

\bibitem{Lowe2005} Lowe, R. J.; Koseff, J. R.; Monismith, S. G. Oscillatory
flow through submerged canopies: 1. Velocity structure. \textit{J. Geophys.
Res. Oceans} \textbf{2005}, \textit{110(C10)}, C10016.
\url{https://doi.org/10.1029/2004JC002788}

\bibitem{MadinConnolly2006} Madin, J. S.; Connolly, S. R. Ecological
consequences of major hydrodynamic disturbances on coral reefs.
\textit{Nature} \textbf{2006}, \textit{444}, 477--480.
\url{https://doi.org/10.1038/nature05328}

\bibitem{McManus1997} McManus, J. W.; Reyes, R. B.; Na\~{n}ola, C. L.
Effects of Some Destructive Fishing Methods on Coral Cover and Potential Rates of Recovery. \textit{Environ. Manage.} \textbf{1997}, \textit{21},
69--78.
\url{https://doi.org/10.1007/s002679900006}

\bibitem{Meyers1994} Meyers, M. A. \textit{Dynamic Behavior of Materials};
Wiley: New York, NY, USA, 1994.
\url{https://doi.org/10.1002/9780470172278}

\bibitem{Minnaert1933} Minnaert, M. On musical air-bubbles and the sounds of
running water. \textit{London Edinburgh Dublin Philos. Mag. J. Sci.}
\textbf{1933}, \textit{16(104)}, 235--248.
\url{https://doi.org/10.1080/14786443309462277}

\bibitem{Monismith2007} Monismith, S. G. Hydrodynamics of Coral Reefs.
\textit{Annu. Rev. Fluid Mech.} \textbf{2007}, \textit{39}, 37--55.
\url{https://doi.org/10.1146/annurev.fluid.38.050304.092125}

\bibitem{NayfehMook1979} Nayfeh, A. H.; Mook, D. T. \textit{Nonlinear
Oscillations}; Wiley: New York, NY, USA, 1995.
\url{https://doi.org/10.1002/9783527617586}

\bibitem{Nepf2012} Nepf, H. M. Flow and Transport in Regions with Aquatic
Vegetation. \textit{Annu. Rev. Fluid Mech.} \textbf{2012}, \textit{44},
123--142.
\url{https://doi.org/10.1146/annurev-fluid-120710-101048}

\bibitem{Noordzij1974} Noordzij, L.; van Wijngaarden, L. Relaxation effects,
caused by relative motion, on shock waves in gas-bubble/liquid mixtures.
\textit{J. Fluid Mech.} \textbf{1974}, \textit{66(1)}, 115--143.
\url{https://doi.org/10.1017/S0022112074000103}

\bibitem{PetSoede1999} Pet-Soede, C.; Cesar, H. S. J.; Pet, J. S. An
economic analysis of blast fishing on Indonesian coral reefs.
\textit{Environ. Conserv.} \textbf{1999}, \textit{26(2)}, 83--93.
\url{https://doi.org/10.1017/S0376892999000132}

\bibitem{Pierce2019} Pierce, A. D. \textit{Acoustics: An Introduction to Its
Physical Principles and Applications}, 3rd ed.; Springer: Cham,
Switzerland, 2019.
\url{https://doi.org/10.1007/978-3-030-11214-1}

\bibitem{PlessetProsperetti1977} Plesset, M. S.; Prosperetti, A. Bubble
dynamics and cavitation. \textit{Annu. Rev. Fluid Mech.} \textbf{1977},
\textit{9}, 145--185.
\url{https://doi.org/10.1146/annurev.fl.09.010177.001045}

\bibitem{Saila1993} Saila, S. B.; Kocic, V. Lj.; McManus, J. W. Modelling
the effects of destructive fishing practices on tropical coral reefs.
\textit{Mar. Ecol. Prog. Ser.} \textbf{1993}, \textit{94(1)}, 51--60.
\url{https://doi.org/10.3354/meps094051}

\bibitem{Showen2018} Showen, R.; Dunson, C.; Woodman, G. H.;
Christopher, S.; Lim, T.; Wilson, S. C. Locating fish bomb blasts in
real-time using a networked acoustic system. \textit{Mar. Pollut. Bull.}
\textbf{2018}, \textit{128}, 496--507.
\url{https://doi.org/10.1016/j.marpolbul.2018.01.029}

\bibitem{SolowayDahl2014} Soloway, A. G.; Dahl, P. H. Peak sound pressure
and sound exposure level from underwater explosions in shallow water.
\textit{J. Acoust. Soc. Am.} \textbf{2014}, \textit{136(3)}, EL218--EL223. \url{https://doi.org/10.1121/1.4892668}

\bibitem{Thomson1950} Thomson, W. T. Transmission of Elastic Waves through a Stratified Solid Medium. \textit{J. Appl. Phys.} \textbf{1950},
\textit{21(2)}, 89--93.
\url{https://doi.org/10.1063/1.1699629}

\bibitem{Virieux1986} Virieux, J. P-SV wave propagation in heterogeneous
media: Velocity-stress finite-difference method. \textit{Geophysics}
\textbf{1986}, \textit{51(4)}, 889--901.
\url{https://doi.org/10.1190/1.1442147}

\bibitem{Virtanen2020} Virtanen, P.; Gommers, R.; Oliphant, T. E.;
Haberland, M.; Reddy, T.; Cournapeau, D.; Burovski, E.; Peterson, P.;
Weckesser, W.; Bright, J.; van der Walt, S. J.; Brett, M.; Wilson, J.;
Millman, K. J.; Mayorov, N.; Nelson, A. R. J.; Jones, E.; Kern, R.;
Larson, E.; Carey, C. J.; Polat, {\.I}.; Feng, Y.; Moore, E. W.;
VanderPlas, J.; SciPy 1.0 Contributors. SciPy 1.0: Fundamental
algorithms for scientific computing in Python. \textit{Nat. Methods}
\textbf{2020}, \textit{17}, 261--272.
\url{https://doi.org/10.1038/s41592-019-0686-2}

\bibitem{vanWijngaarden1972} van Wijngaarden, L. One-Dimensional Flow of Liquids Containing Small Gas Bubbles. \textit{Annu. Rev. Fluid Mech.}
\textbf{1972}, \textit{4}, 369--396.
\url{https://doi.org/10.1146/annurev.fl.04.010172.002101}

\bibitem{Wood1930} Wood, A. B. \textit{A Textbook of Sound}; G. Bell and
Sons: London, UK, 1930.

\bibitem{Woodman2003} Woodman, G. H.; Wilson, S. C.; Li, V. Y. F.;
Renneberg, R. Acoustic characteristics of fish bombing: Potential to
develop an automated blast detector. \textit{Mar. Pollut. Bull.}
\textbf{2003}, \textit{46(1)}, 99--106. \url{https://doi.org/10.1016/S0025-326X(02)00322-3}

\bibitem{Yee1966} Yee, K. S. Numerical solution of initial boundary value
problems involving Maxwell's equations in isotropic media. \textit{IEEE
Trans. Antennas Propag.} \textbf{1966}, \textit{14(3)}, 302--307.
\url{https://doi.org/10.1109/TAP.1966.1138693}

\end{thebibliography}
\end{document}